\documentclass[%
 reprint,
 superscriptaddress,
 showpacs,preprintnumbers,
 amsmath,amssymb,
 aps,
 prl,
]{revtex4-1}

\usepackage{graphicx}         
\usepackage{dcolumn}          
\usepackage{bm}               
\usepackage{color}

\begin{document}

\title{Raman Images of a Single Molecule in a Highly Confined
Plasmonic Field}

\author{Sai Duan}
\affiliation{Hefei National Laboratory for Physical Science at the
Microscale and Synergetic Innovation Center of Quantum Information \&
Quantum Physics, University of Science and Technology of China, Hefei,
230026 Anhui, P. R. China.}
\affiliation{Department of Theoretical Chemistry and Biology, School of
Biotechnology, Royal Institute of Technology, S-106 91 Stockholm,
Sweden.}

\author{Guangjun Tian}%
\affiliation{Department of Theoretical Chemistry and Biology, School of
Biotechnology, Royal Institute of Technology, S-106 91 Stockholm,
Sweden.}

\author{Yongfei Ji}%
\affiliation{Department of Theoretical Chemistry and Biology, School of
Biotechnology, Royal Institute of Technology, S-106 91 Stockholm,
Sweden.}

\author{Jiushu Shao}
\affiliation{Key Laboratory of Theoretical Computational Photochemistry,
Ministry of Education, College of Chemistry, Beijing Normal University,
Beijing 100875, P. R. China.}%

\author{Zhenchao Dong}
\affiliation{Hefei National Laboratory for Physical Science at the
Microscale and Synergetic Innovation Center of Quantum Information \&
Quantum Physics, University of Science and Technology of China, Hefei,
230026 Anhui, P. R. China.}

\author{Yi Luo}
\email{yiluo@ustc.edu.cn}
\affiliation{Hefei National Laboratory for Physical Science at the
Microscale and Synergetic Innovation Center of Quantum Information \&
Quantum Physics, University of Science and Technology of China, Hefei,
230026 Anhui, P. R. China.}
\affiliation{Department of Theoretical Chemistry and Biology, School of
Biotechnology, Royal Institute of Technology, S-106 91 Stockholm,
Sweden.}%

\date{\today}

\begin{abstract}
Under the local plasmonic excitation, the Raman images of a single
molecule can now reach sub-nanometer resolution. We report here a
theoretical description of the interaction between a molecule and
a highly confined plasmonic field. It is shown that when the spatial
distribution of the plasmonic field is comparable with the size of
the molecule, the optical transition matrix of the molecule
becomes to be dependent on the position and the
spatial distribution of the plasmonic field, resulting in spatially
resolved Raman image of a molecule. It is found that the resonant Raman
image reflects the electronic transition density of the molecule.  In
combination with the first principles calculations, the simulated Raman
image of a porphyrin derivative adsorbed on the silver surface nicely
reproduces its experimental counterpart. The present theory provides the
basic framework for describing linear and nonlinear responses of
molecules under the highly confined plasmonic field.
\end{abstract}



\maketitle
The development of tip enhanced Raman scattering (TERS) technique has
significantly increased the spatial resolution of Raman images for
molecules\cite{ters1,ters2,zhang2013nature}. Under the low temperature
and ultrahigh vacuum conditions, the resolution has amazingly reached a
sub-nanometer level for a porphyrin derivative adsorbed on the silver
surface\cite{zhang2013nature}. It is anticipated that the spatial
confinement of the tip-induced plasmon has played a decisive role in
achieving such a high resolution. In this case, the spatial distribution
of the plasmonic field has to be comparable with the size of the
molecule, even with the inclusion of possible nonlinear
processes\cite{zhang2013nature}. This situation presents a great
challenge to the conventional theory, which always assumes that the
electromagnetic (EM) field uniformly interacts with the
molecule\cite{long2002}. A new theory that takes into account the
locality of the EM thus needs to be developed.  Moreover, what a Raman
image really tells about the molecular structure is another important
issue that has not yet been discussed in the literature.

In this work, we have derived a theoretical framework to describe
Raman images of a molecule as observed in the TERS experiments.  
In combination with the first principles calculations, we have 
successfully reproduced experimental Raman image
of the porphyrin derivative adsorbed on the silver surface. It is found
that within the Born-Oppenheimer approximation, the resonant Raman image
reflects the density of the electronic transition between the ground and
the excited states.  The role of linear and nonlinear processes on the
resolution of the Raman images has been identified.

In the TERS experiments, a nano-cavity formed in-between the tip and the
substrate is the host of the spatially confined plasmon (SCP). For the
metals used in the experiments, the plasmonic frequency often falls into
the visible or ultraviolet-visible region, hence the electric dipole
approximation\cite{scully1997} can still hold. The interaction
Hamiltonian between SCP and adsorbates could be described
by\cite{scully1997}
\begin{equation}
\mathcal{H}^\prime=-e\hat{\mathbf{r}}\cdot\hat{\bm{E}}(\mathbf{r},\bm{R}_T,t),
\label{eq:int}
\end{equation}
where $e$ is the elementary charge, $\hat{\mathbf{r}}$ is the electron
position operator and $\hat{\bm{E}}$ is the operator for electric field
of SCP which is obviously related to the tip position $\bm{R}_T$. In
this case, $\hat{\bm{E}}$ cannot be treated uniformly in space due to
its specific spatial distribution. Consequently, the optical transition
matrix element between two states is determined by
$\langle\psi_g|\hat{\mathbf{r}}\hat{\bm{E}}(\mathbf{r},\bm{R}_T)|\psi_r\rangle$
rather than $\langle\psi_g|\hat{\mathbf{r}}|\psi_r\rangle$ as in the
conventional response theory\cite{albrecht1961jcp,lee2004jcp}. This
modification implies that the optical processes would be dependent on
the position of the tip that hosts the SCP. This can naturally explain
why it is possible to obtain super-high spatial resolution of the Raman
images.

By taking into account the position dependent electric field, we can
re-derive the expression for the spontaneous resonant linear Raman
processes following the Albrecht's
theory\cite{albrecht1961jcp,long2002}. The amplitude of induced linear
polarization could be calculated as the summation of Franck-Condon (FC,
$A$) and Herzberg-Teller (HT, $B$) terms,
\begin{equation}
\bm{P}_0^L=\bm{P}_{0,A}^L+\bm{P}_{0,B}^L,
\label{eq:pl}
\end{equation}
where
\begin{equation}
\begin{split}
\bm{P}_{0,A}^L&=\frac{\sqrt{F_P}M_iE_i^0}{\hbar}
\Bigg[\langle\psi_g|\hat{\mathbf{r}}|\psi_r\rangle
\langle\psi_r|\hat{\mathbf{r}}g(\mathbf{r},\bm{R}_T)|\psi_g\rangle\\
&\times\sum_{v^r=0}^\infty
\frac{\langle{v}^f|v^r\rangle\langle{v}^r|v^i\rangle}
{\omega_{e^rv^r:e^gv^i}-\omega_p-\imath\Gamma}\Bigg]+\textrm{NRT}
\end{split}
\label{eq:FCL}
\end{equation}
\begin{equation}
\begin{split}
\bm{P}_{0,B}^L&=\frac{\sqrt{F_P}M_iE_i^0}{\hbar}
\Bigg[\frac{\partial\langle\psi_g|\hat{\mathbf{r}}|\psi_r\rangle}{\partial{Q}_k}
\langle\psi_r|\hat{\mathbf{r}}g(\mathbf{r},\bm{R}_T)|\psi_g\rangle\\
&\times\sum_{v^r=0}^\infty
\frac{\langle{v}^f|Q_k|v^r\rangle\langle{v}^r|v^i\rangle}
{\omega_{e^rv^r:e^gv^i}-\omega_p-\imath\Gamma}\\
&+\langle\psi_g|\hat{\mathbf{r}}|\psi_r\rangle
\frac{\partial\langle\psi_r|\hat{\mathbf{r}}g(\mathbf{r},\bm{R}_T)|\psi_g\rangle}{\partial{Q}_k}\\
&\times\sum_{v^r=0}^\infty
\frac{\langle{v}^f|v^r\rangle\langle{v}^r|Q_k|v^i\rangle}
{\omega_{e^rv^r:e^gv^i}-\omega_p-\imath\Gamma}\Bigg]+\textrm{NRT}.
\end{split}
\label{eq:HTL}
\end{equation}
Here $F_P$ is the Purcell factor\cite{purcell1946pr} which accounts for
the enhancement of spontaneous emission in nano-cavity and is
independent on the position of the TERS tip in the $xy$ plane, $E_i^0$
is the electric field amplitude of incident laser, $M_i$ is the
enhancement factor of the SCP with respect to the incident laser,
$\hbar$ is the reduced Planck constant, $|\psi_g\rangle$ and
$|\psi_r\rangle$ are the electronic ground and resonant excited states,
$g$ is the corresponding distribution function of the electric field
amplitude of the SCP with proper normalization, $Q_k$ is the
corresponding normal mode, $|v^i\rangle$ and $|v^f\rangle$ are the
initial and final vibrational states in $|\psi_g\rangle$, $|v^r\rangle$
is the vibrational state in $|\psi_r\rangle$, $\omega_{e^rv^r:e^gv^i}$
is the frequency difference between $|\psi_r\rangle|v^r\rangle$ and
$|\psi_g\rangle|v^i\rangle$, $\omega_p$ is the frequency of the plasmon
generated by the incident light, $\Gamma$ is the damping factor, and
$\textrm{NRT}$ is the non-resonant term. From the polarization, the
Raman intensity can be calculated directly
as\cite{long2002,albrecht1961jcp,jcc-raman}
\begin{equation}
I_{s}=\frac{\pi^2c\tilde{\nu}_s^4M_d^2|\bm{P}_0^L|^2}{2\epsilon_0},
\label{eq:Is}
\end{equation}
where $c$ is the speed of light, $\tilde{\nu}_s$ is the wave number of
scattering, $\epsilon_0$ is the permittivity of free space, and $M_d$ is
the directional radiation pattern factor\cite{leru2009}. Here the total
polarization was treated as a classical oscillating
dipole\cite{jcc-raman}. We have noted that the present consideration of
SCP is equivalent to the quantization scheme proposed by Archambault
\textit{et al.}\cite{archambault2010prb} for the propagation of the
surface plasmonic waves as well as the classical treatment proposed by
Xu and co-workers\cite{xu2004prl,johansson2005prb}. The key difference
here is to consider the effects of the distribution function $g$ 
to the optical transition matrix. It should also be mentioned that 
the first order Taylor expansion for $g(\mathbf{r},\bm{R}_T)$ would 
naturally account for the electric field gradient effects in 
Raman spectroscopy as discussed in the 
literature\cite{sass1981jpc,ayars2000prl,nobusada2009pra}.

\begin{figure*}
\centering
\includegraphics[width=0.75\textwidth]{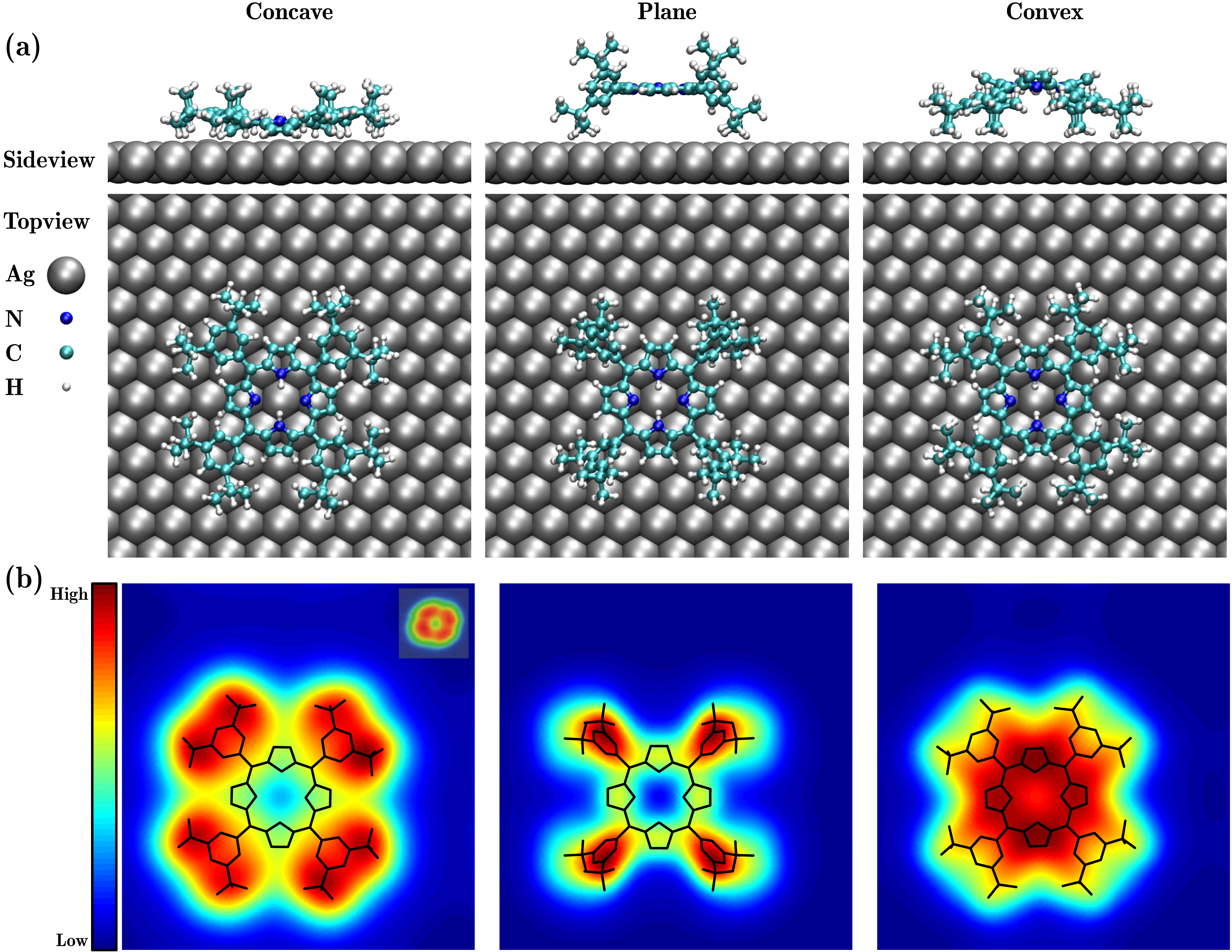}
\caption{(a) Optimized structures of one H$_2$TBPP tautomer adsorbed on
the Ag(111) surface for concave, plane, and convex configurations,
respectively.  Gray, blue, cyan and white balls represent Ag, N, C, and
H atoms, respectively. Only the topmost slab layer of the Ag(111)
surface as well as all atoms of adsorbates in one supercell are depicted
for clarity.  (b) Calculated average STM images with the sample biased
by 1.0~V for concave, plane, and convex configurations, respectively.
The solid lines represent the skeleton of H$_2$TBPP. The insert figure
shows the experimental STM image under the same condition extracted from
Ref.~\citenum{zhang2013nature}.}
\label{fig:mol}
\end{figure*}

For the sake of computations, we have chosen to expand the function $g$  
in terms of the Gaussian basis sets
\begin{equation}
g(\mathbf{r},\bm{R}_T)=\sum_{D}\sum_{l,m,n}\sum_{\alpha}c_{\alpha,D}^{lmn}g_{\alpha,D}^{lmn},
\label{eq:n-ex}
\end{equation}
where $g_{\alpha,D}^{lmn}$ is a Gaussian function localized at the
center $\mathbf{r}_D$ with exponent $\alpha$, which can be written as
\begin{equation}
g_{\alpha,D}^{lmn}=(x-x_D)^l(y-y_D)^m(z-z_D)^ne^{-\alpha(\mathbf{r}-\mathbf{r}_D)^2}
\label{eq:gd}
\end{equation}
and $c_{\alpha,D}^{lmn}$ is the corresponding coefficient. Here
$\mathbf{r}_D$ may represent the position of the SCP, which could be
obtained by fitting the realistic electric field distribution and in
principle it is not necessarily equal to $\bm{R}_T$. As the first
demonstration, only the $s$-type Gaussian functions are considered for
$g_{\alpha,D}^{lmn}$. It is noted that for absolute Raman intensities
$g$ should satisfy some proper normalization conditions. However, for
the Raman images, the relative values are adequate. For the practical
calculations, we chose $c_{\alpha,D}^{lmn}=1$ and $l=m=n=0$ in
Eq.~\ref{eq:n-ex}.  Moreover, for a molecule under the TERS tip, only
the $zz$ component needs to be evaluated\cite{zhang2013nature}.

We put the new theory to the test by directly simulating the system that
was measured in a recent study\cite{zhang2013nature}, i.e. a single
\textit{meso}-tetrakis-(3,5-di-tertiarybutylphenyl)-porphyrin
(H$_2$TBPP) molecule adsorbed on the silver (Ag) surface. The high
resolution scanning tunneling microscope (STM) and Raman
images\cite{zhang2013nature} provide good references for theoretical
modeling. The details of the density functional theory calculations are
given in the Supplemental Material\cite{ms}.

Here we considered two degenerate tautomers of H$_2$TBPP
molecule\cite{qfu2009apl}. The optimized structures of one H$_2$TBPP
tautomer adsorbed on the Ag(111) surface are depicted in
Fig.~\ref{fig:mol}(a)\cite{ms}. The optimized structures of the other
tautomers are similar except the central hydrogens bonded to different
nitrogen atoms. We have considered three configurations: concave, plane,
and convex, respectively. The first and last configurations were
identified when H$_2$TBPP adsorbed on the Cu(111) surface by
STM\cite{ditze2014jacs} and the second configuration is the minimum of
the isolated molecule. Our calculations have shown that the concave is
the most stable adsorption configuration, owing to the long range
dispersion included in current calculations\cite{grimme2006jcc}.
Meanwhile, the second stable configuration is the plane and the convex
is the least stable one. The calculated STM images\cite{ms} of all
configurations, together with the experimental
result\cite{zhang2013nature} are given in Fig.~\ref{fig:mol}(b).  One
can immediately see that only the calculated STM image of the concave
resembles well the experimental image. It can thus be concluded that
H$_2$TBPP adsorbed on the Ag(111) surface has the concave configuration
under the experimental conditions.

\begin{figure}
\centering
\includegraphics[width=0.48\textwidth]{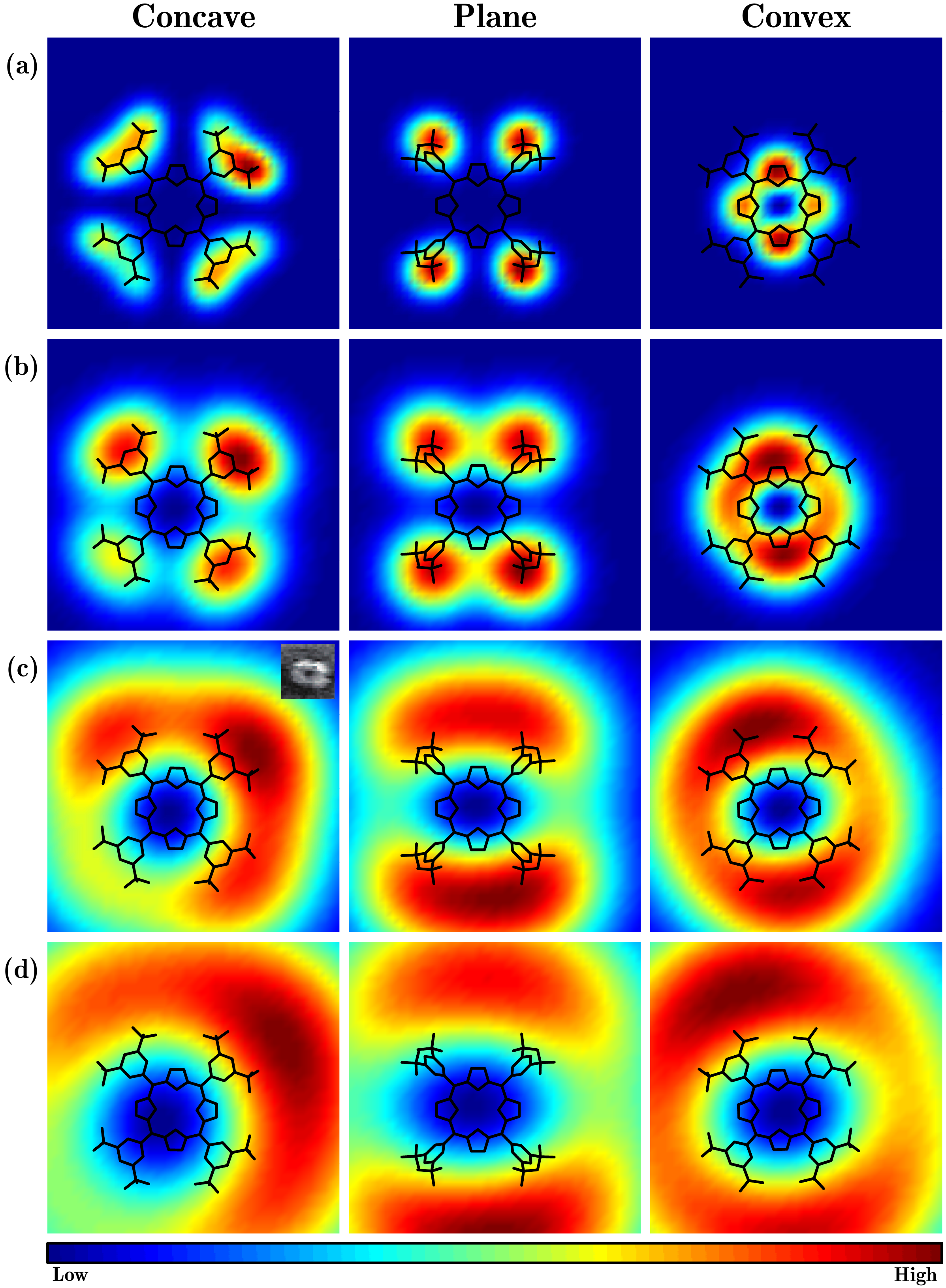}
\caption{Calculated linear Raman images from the FC term with the $x$
and $y$ components of full width at half maximum to be (a) 5~\AA{}, (b)
10~\AA{}, (c) 20~\AA{}, and (d) 30~\AA{} for concave, plane, and convex
configurations, respectively. The solid lines represent the skeleton of
H$_2$TBPP, while, the insert figure shows the experimental Raman image
extracted from Ref.~\citenum{zhang2013nature}.}
\label{fig:izz}
\end{figure}

To simulate the Raman image, H$_2$TBPP was extracted from the optimized
adsorption structures and its excited states were calculated by the time
dependent density functional theory at the hybrid B3LYP level with
6-31G(d) basis sets\cite{ms}.  It should be mentioned that it is the
compromise that we have to make for such large systems from
computational point of view. It is also a reasonable
approximation since the molecule is only physisorbed on the Ag(111)
surface. The strong absorption states for three configurations are found
to be 2.47, 2.36, and 2.25~eV, respectively, in the region of the
excitation energy of the experiment (532~nm,2.33~eV)\cite{zhang2013nature}. 
We have thus chosen these three excited states to simulate the resonant 
Raman images. Consistent with the experimental setup\cite{zhang2013nature,ho2003science}, 
the center of SCP, i.e. $\mathbf{r}_D$, is in the plane which is 
about 2~\AA{} above the highest position of the adsorbates.  
Meanwhile, along the $x$ and $y$ directions, the full width at half 
maximum (FWHM) of the plasmonic field, $g_{\alpha,D}^{000}$, was set 
to be 5, 10, 20, and 30~\AA{}, respectively, while the $z$ component 
was fixed at 5~\AA{}. We should emphasize that the calculated Raman 
images are not sensitive to small changes of $\mathbf{r}_{D}$ and 
the $\alpha$ in the $z$ component of $g_{\alpha,D}^{000}$.

Under the resonant condition, for allowed transitions, the FC term
$P_{0,A}^L$ is known to be the dominant one\cite{myers2008jpca}. All
simulated linear Raman images from the FC term are presented in
Fig.~\ref{fig:izz}. It can be seen that the size of the Raman image is
dependent on the size of the SCP. This implies that the precondition for
the high resolution Raman image is to generate a highly focused
plasmonic field. It is nice to observe that different configuration of
the molecule gives very different Raman image, indicating that TERS
technique is a powerful tool to study the structure of adsorbates. One
can notice that the Raman image of the concave with FWHM of 20~\AA{} is
in very good agreement with the experimental image, which is consistent
with the energy and STM calculations. Here the symmetry breaking of the
calculated Raman images should be attributed to the interaction between
adsorbates and the Ag(111) substrate. By definition, Raman images
reflect the density change involved in the electronic transition, rather
than the local density of state of the adsorbate. This is naturally
reflected by the obvious difference between the Raman and STM images.

\begin{figure}
\centering
\includegraphics[width=0.48\textwidth]{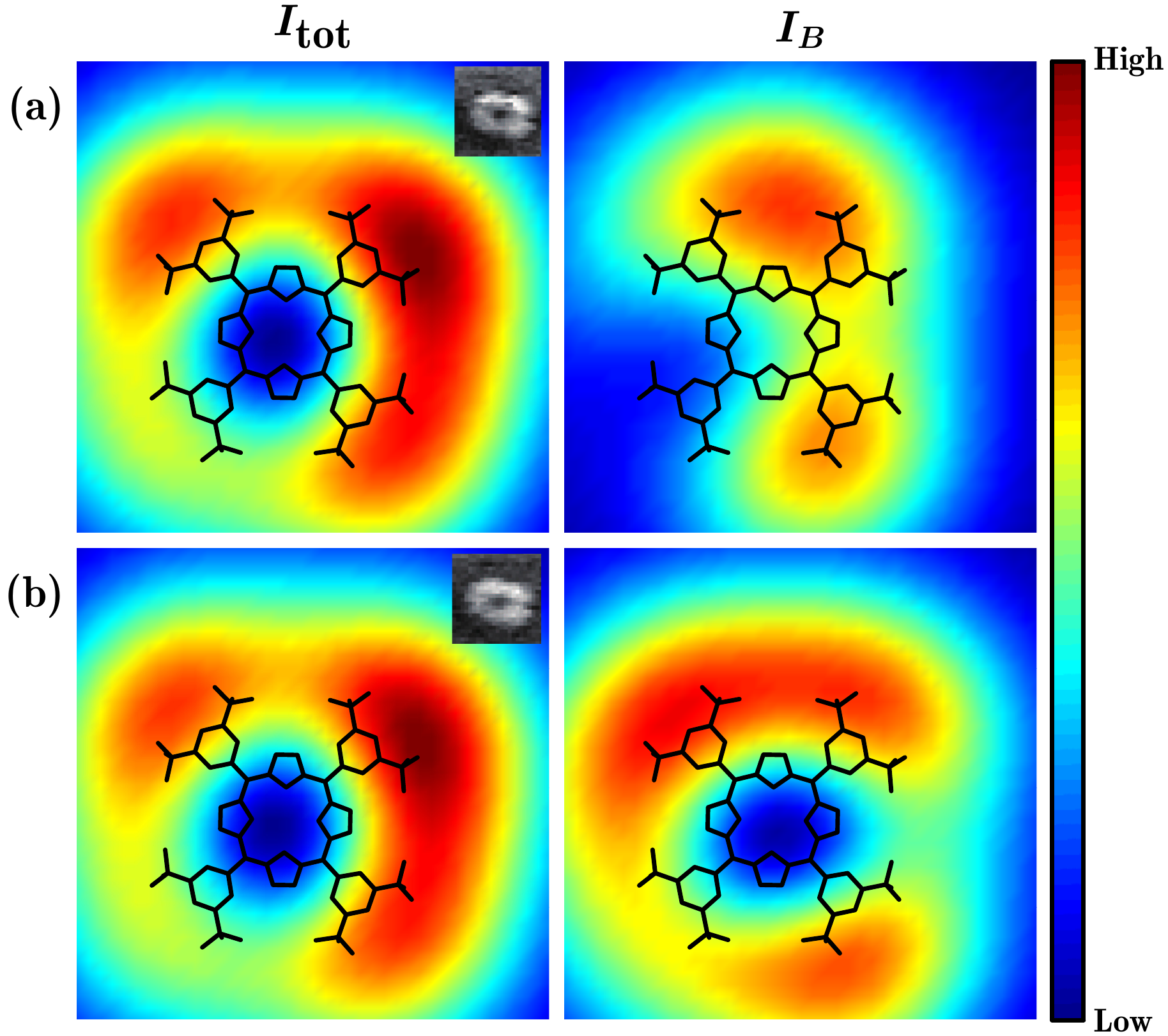}
\caption{Calculated linear Raman images of two bands at (a)
820~cm$^{-1}$ and (b) 1200~cm$^{-1}$ from total polarization
($I_{\textrm{tot}}$) and only the HT term ($I_B$) with the full width at
half maximum of 20~\AA{} for the concave configuration. The $I_B$ was
scaled by a factor of 100 and 400 for 800 and 1200~cm$^{-1}$,
respectively. The solid lines represent the skeleton of H$_2$TBPP,
while, the insert figures are the corresponding experimental Raman
images extracted from Ref.~\citenum{zhang2013nature}.}
\label{fig:bterm}
\end{figure}

The HT term $P_{0,B}^L$ is dependent on the vibrational modes and its
contribution can thus be used to identify the vibrations of the
molecule.  We have evaluated numerically the effect of the $B$ term for
two vibrational bands, around 820 and 1200~cm$^{-1}$, by using the
linear coupling model\cite{macak2000cpl}. The calculated Raman images
from the total polarization as well as the HT term alone for these two
bands are shown in Fig.~\ref{fig:bterm}.  One can immediately notice
that the HT term is very sensitive to the vibrational modes. Its
contribution to the total intensity holds the key to distinguish Raman
images of different vibrations. For the H$_2$TBPP molecule, the
contribution of the HT term is very small. Hence, the Raman images from
total intensity calculations appear to be identical for the two
vibrational bands. It should be mentioned that the Raman calculations
are performed for the single molecule. The inclusion of the substrate
could increase the contribution of the HT term. This could be a reason
behind the small variation observed in the experimental Raman
images\cite{zhang2013nature}.  We should emphasize that the calculated
Raman images given in Ref.~\citenum{zhang2013nature} were obtained by
assuming that the confined plasmonic field does not alter the transition
matrix itself, which is a simplified approximation to the basic theory
presented in this work.

It was found experimentally that the observed Raman intensity was
nonlinearly dependent on the power of the incident
light\cite{zhang2013nature}. The contributions of the linear and
nonlinear processes to the total intensity were found to be 40\% and
60\%, respectively, at the saturation condition\cite{zhang2013nature}.
It is thus necessary to examine how the nonlinear process affects the
Raman images of the molecule. Three processes, namely stimulated Raman
as well as two hot luminescence processes (I and II), could contribute
to the nonlinear Raman signal, when both pump and broadband SCPs are
involved. In analogy to the theory proposed by Lee \textit{et
al.}\cite{lee2004jcp}, the amplitude of induced nonlinear polarization
also consists two terms
\begin{equation}
\bm{P}_0^{NL}=\bm{P}_{0,A}^{NL}+\bm{P}_{0,B}^{NL},
\label{eq:pnl}
\end{equation}
where
\begin{equation}
\begin{split}
\bm{P}_{0,A}^{NL}
&=\frac{M_i^2M_s|E_i^0|^3\sqrt{2\pi}\tau_s}{\hbar^3}
\langle\psi_r|\hat{\mathbf{r}}|\psi_g\rangle
\langle\psi_r|\hat{\mathbf{r}}g(\mathbf{r},\bm{R}_T)|\psi_g\rangle^3\\
&\times\sum_{v^r,v^{r\prime}}
\langle{v^f}|v^r\rangle\langle{v^r}|v^i\rangle\langle{v^i}|v^{r\prime}\rangle\langle{v}^{r\prime}|v^f\rangle\lambda^{(3)}
+\textrm{NRT}.
\end{split}
\label{eq:FCNL}
\end{equation}
Here $M_s$ is the enhancement factor of the broadband SCP that has the
duration of $2\sqrt{2\ln2}\tau_s$. The expression of $\bm{P}_{0,B}^{NL}$
and the definition of $\lambda^{(3)}$ could be found in Supplemental
Material\cite{ms}. We would emphasize that $\lambda^{(3)}$ is
independent on the position of the TERS tip.

\begin{figure}
\centering
\includegraphics[width=0.48\textwidth]{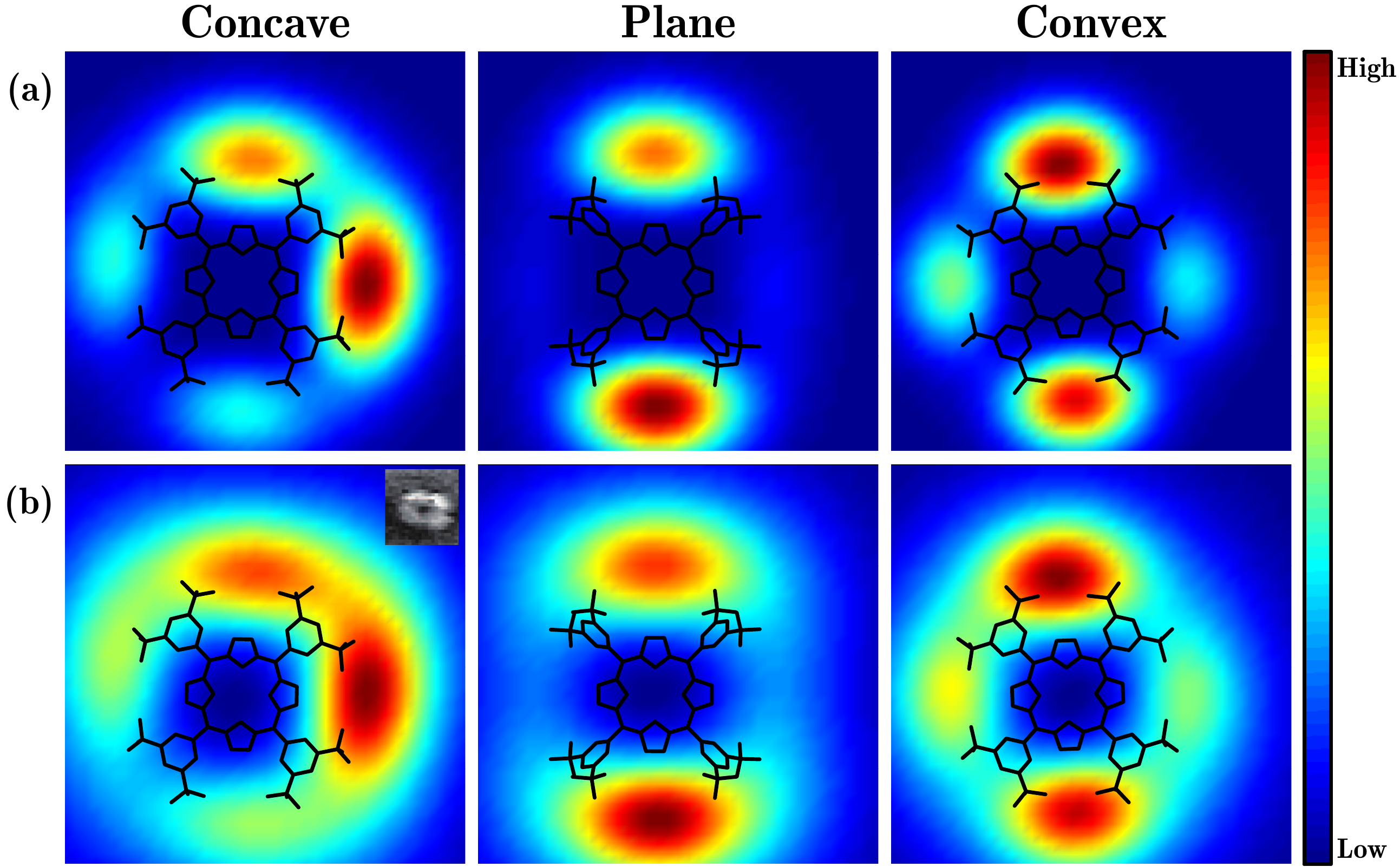}
\caption{Calculated (a) nonlinear and (b) total Raman images for
concave, plane, and convex configurations, respectively, under the
plasmonic field with the full width at half maximum of 20~\AA{}.  The
solid lines represent the skeleton of H$_2$TBPP, while, the insert
figure shows the experimental Raman image extracted from
Ref.~\citenum{zhang2013nature}.}
\label{fig:inzz}
\end{figure}

Similar to the linear process, the FC term is also completely dominant
in the nonlinear process. Once the $\bm{P}_0^{NL}$ is obtained,
$|\bm{P}_0^L|^2$ in Eq.~\ref{eq:Is} could be replaced by
$|\bm{P}_0^{NL}|^2$ or their summation for nonlinear or total Raman
intensities, respectively, because the phases of them are
unrelated\cite{chemla1980rpp}.  Admittedly, obtaining the absolute value
of linear and nonlinear polarizations requires the solution of the
classical Maxwell's equations for realistic nanostructures.  Instead, we
use the experimentally determined contributions of the linear and
nonlinear terms in the saturation condition\cite{zhang2013nature} to
define the pre-factors. The calculated nonlinear Raman images with FWHM
of 20~\AA{} are depicted in Fig.~\ref{fig:inzz}(a).  In general, the
nonlinear images do have higher spatial resolution than their linear
counterparts. However, for this particular system, the improvement is
not as much as one anticipated\cite{zhang2013nature}.  The calculated
total Raman image (including both linear and nonlinear effects) from the
concave configuration with the field distribution size of 20~\AA{},
gives the best agreement with the experimental image as shown in
Fig.~\ref{fig:inzz}(b). This result reveals that the actual size of the
experimentally confined plasmonic field could be close to 20~\AA{},
which is consistent with our numerical EM simulations\cite{ms}
and may be the decisive factor for the sub-nanometer Raman images.
However, the increase of the nonlinear contribution to the total
intensity will on the other hand be an effective way to further improve
the resolution of the Raman image.

In summary, we have proposed a quantum mechanical description for the
interaction between a molecule and a highly confined plasmonic field. It
shows that the SCP could modify the transition matrix and result in the
Raman images with high resolution. The usefulness of the description is
highlighted by reproducing successfully the experimental Raman images of
a H$_2$TBPP molecule adsorbed on the Ag(111) surface. The theoretical
framework established in this work lays the foundation for the future
development of linear and nonlinear plasmonic spectroscopy.

This work was supported by the Ministry of Science and Technology of China
(2010CB923300), Natural Science Foundation of China (21121003),
Strategic Priority Research Program of Chinese Academy of
Sciences (XDB01020200), G{\"o}ran Gustafsson Foundation for Research in
Natural Sciences and Medicine, and the Swedish Research Council (VR).  The
Swedish National Infrastructure for Computing (SNIC) was acknowledged
for computer time.

%

\end{document}